\documentclass[doublecol,linenumbers]{epl2}
\usepackage{latexsym}
\usepackage{graphicx}
\usepackage{amssymb}
\usepackage{amsmath}
\usepackage{epstopdf}
\usepackage{float}
\usepackage{ulem}
\usepackage{subeqnarray}
\title{Viscoelastic effects and anomalous transient levelling exponents  in thin films}
\shorttitle{Viscoelastic effects  and anomalous transient levelling exponents in thin films} 
\author{M. Benzaquen\inst{1} \and T. Salez\inst{1} \and E. Rapha\"el\inst{1}}
\shortauthor{M. Benzaquen \etal}
\institute{                    
  \inst{1} Laboratoire de Physico-Chimie Th\'eorique, UMR CNRS 7083 Gulliver - ESPCI ParisTech, 10 rue Vauquelin, 75005, Paris}
\abstract{
We study theoretically the profile evolution of a thin viscoelastic film supported onto a no-slip flat substrate. Due to the nonconstant initial curvature at the free surface, there is a flow driven by Laplace pressure and mediated by viscoelasticity. In the framework of lubrication theory, we derive a thin film equation that contains local viscoelastic stress through the Maxwell model. Then, considering a {sufficiently regular} small perturbation of the free surface, we linearise the equation and  {derive its general solution}. We analyse and discuss in details the behaviour of this function. We then use it to study the viscoelastic evolution of a  Gaussian initial perturbation through its transient levelling exponent. For initial widths of the profile that are smaller than a characteristic length scale {involving} both the film thickness and the elastocapillary length, this {exponent} is shown to reach anomalously high values at the elastic-to-viscous transition. This prediction should in particular be observed  at sufficiently short times in experiments on thin polymer films.
}
\begin{document}
\maketitle

\begin{nolinenumbers}
Over the past decades, the study of soft materials in confined geometries and thin films \cite{Oron1997,Craster2009,Blossey2012} has widely attracted the interest  of physicists, biophysicists, chemists and engineers. In particular, thin polymer films are nowadays of major importance in several industrial applications such as  biocompatible coatings, organic microelectronics and polymeric data storage devices. In order to gain insights into the behaviour of these films and their constitutive macromolecules, a wide class of experiments, including dewetting \cite{Baumchen2009,Ziebert2009}, nanoindentation with gold particles \cite{Fakhraai2008}, and levelling of stepped films \cite{McGraw2011,McGraw2012,Chai2014}, have been performed.  
Enhanced mobility effects in ultra-thin polymer films  have been predicted \cite{Brochard2000,Si2005}, and observed \cite{Jones1999,Bodiguel2006,Shin2007}. Film preparation by spincoating has also been widely studied \cite{Stillwagon1990,Reiter2001,Raegen2010,Barbero2009}, and is known to govern surface instabilities and pattern formation \cite{Mukherjee2011,Closa2011,Amarandei2012}. 

The evolution of the free surface of a thin Newtonian liquid film with nonconstant curvature is driven by the Laplace pressure and mediated by viscosity. This is well understood from the theoretical point of view  through the so-called capillary-driven thin film equation \cite{Orchard1961,Oron1997,Craster2009,Blossey2012}. 
{Extensive analytical work  on the thin film equation \cite{Myers1998,Bowen2006,Bernis1990,Kondic2003,Salez2012a}  and numerous accurate numerical schemes \cite{Bertozzi1998,Zhornitskaya2000,Sharma1998,Salez2012b} have been performed in the past decades, and have allowed for a deeper understanding of its mathematical features.}
{Long-term traveling-wave solutions have been discussed \cite{Boatto1993}}. 
Convergence of the solutions to intermediate asymptotic regimes \cite{Barenblatt1996} has {also} been revealed. In particular, it was shown that the vertically-rescaled solution for any summable initial profile uniformly converges in time towards a universal self-similar attractor that is precisely given by the Green's function of the capillary-driven linear thin film equation multiplied by the initial algebraic volume of the perturbation \cite{Benzaquen2013}.

The aforementioned capillary-driven thin film equation describes the evolution of thin Newtonian films. Yet, thin films are often made of polymer melts which usually display viscoelastic properties \cite{Macosko1994,Larson1999} in certain temporal ranges. In this case, some parts of the evolution may be different from that of a pure viscous fluid, as the deformation of the system in response to a given stress is now mediated by both {viscosity and elasticity}. The general understanding of viscoelasticity is of great interest in soft condensed matter and physics of glassy systems, as shown by the numerous recent results in the literature. New ways to probe the dynamics of viscoelastic materials have been explored \cite{Pelton2013}. Thin viscoelastic films have been studied experimentally by dewetting \cite{Herminghaus2003,Bodiguel2007}, and relaxation of nanoimprinted patterns \cite{Ding2008,Rognin2012}, in particular. From the theoretical point of view, viscoelastic thin film equations have been derived, using the linear Jeffreys model { \cite{Keunings1987,Rauscher2005,Munch2006,Blossey2006,Blossey2012}}.

In the present work, we {derive} a viscoelastic thin film equation based on the Maxwell model, before linearising it and solving it analytically. In particular, we show that the regular part of this essential solution converges in time towards the Green's function of the purely viscous linear thin film equation \cite{Benzaquen2013}.  Using these  results, we then analyse the temporal  relaxation of a canonical Gaussian height perturbation on a thin viscoelastic film, and study its evolution in terms of a transient levelling exponent \cite{Christov2012}. Finally, we discuss the different observed regimes. We show in particular that the system displays anomalous -- \textit{i.e.} not comprised between the elastic and viscous standard values -- transient levelling exponents for profile widths smaller than a characteristic length scale that is set by the film thickness and the elastocapillary length.
 
\begin{figure}[t!]
\begin{center}
\includegraphics[width= 0.63 \columnwidth]{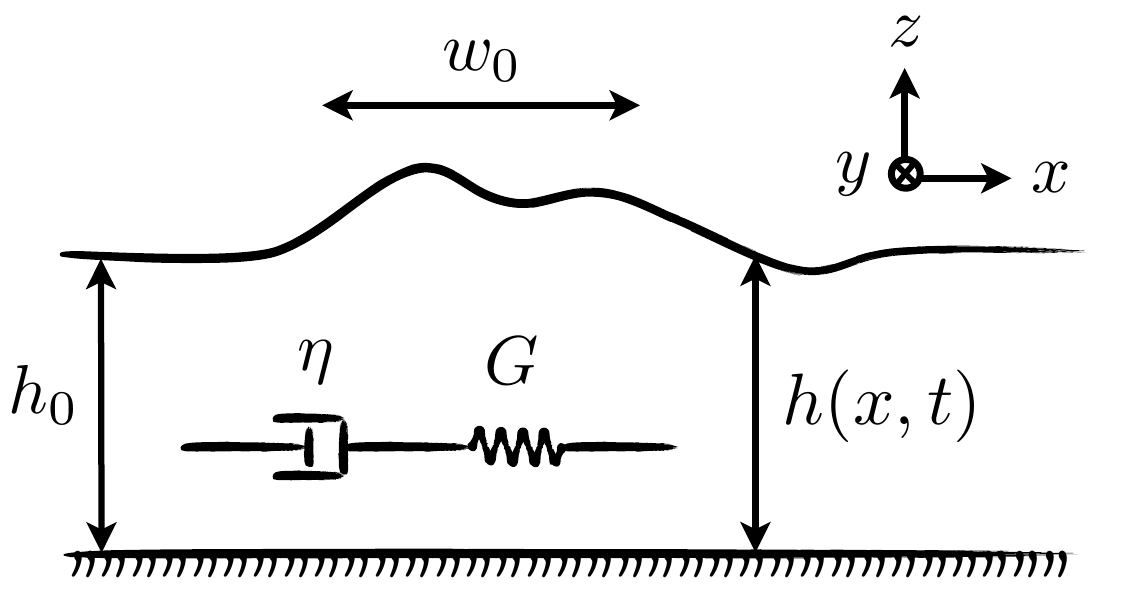}
\end{center}
\caption{Schematic of the vertical height profile $z=h(x,t)$, at horizontal position $x$ and time $t$, of a thin viscoelastic film placed atop a flat substrate. The system is assumed to be spatially invariant in the other horizontal direction $y$. The typical horizontal length scale is denoted by $w_0$, and the thickness at infinity by $h_0$. The intern rheology of the viscoelastic fluid is accounted for through a local Maxwell model (see eq.~(\ref{Maxwell})) of shear viscosity $\eta$ and shear elastic modulus $G$, as symbolised by the cartoon showing the damper and spring in series.} 
\label{fig1}
\end{figure}

\section{Viscoelastic thin film equation}
{ { In this section, we derive a Maxwell-based viscoelastic thin film equation in the spirit of the Jeffreys-based scheme proposed in \cite{Rauscher2005,Munch2006,Blossey2006,Blossey2012}. }}
We consider a 2D thin viscoelastic film lying on a flat substrate (see Fig.~\ref{fig1}). Let the origin $O$ be taken at the substrate level, and let $z=h(x,t)$ be the vertical height profile of the film at horizontal position $x$ and time $t$. The system is assumed to be spatially invariant in the other horizontal direction $y$. Let ${\boldsymbol u} =(u_x,u_z)$ be the 2D velocity field in the  film. For an incompressible flow, the Navier-Stokes equation reads \cite{Landau1987}:
\begin{eqnarray}
\rho \frac{\text{d}{\boldsymbol u}}{\text{d}t} &=&-\boldsymbol \nabla p +\boldsymbol \nabla \sigma'   \ ,\label{NS}
\end{eqnarray}
where $\rho$ is the mass density of the material, $p$ is the pressure, $\sigma'$ is the extra-stress tensor\cite{Landau1987}, and  $\frac{\text{d}}{\text{d}t}=  \partial_t +{\boldsymbol u} \cdot  \boldsymbol \nabla$ is the convective derivative.
Let $h_0$ be the thickness  at infinity, $w_0$  be a typical horizontal dimension (see. Fig.~\ref{fig1}), and   $\epsilon=h_0/ w_0$. Within the the lubrication approximation, one has $\epsilon \ll 1$. We define the space and time dimensionless variables through $x={w_0} X  ,\,  z=h_0 Z  ,\,  h= h_0H  ,\,  t=t_0T$,  where $t_0$ is a typical time scale yet to be determined. Incompressibility implies that the dimensionless velocities read $ u_x=v_0U_X$ and  $u_z=  \epsilon \, v_0U_Z$, where $v_0=w_0/t_0$. In order to determine the scaling for pressure and time, we balance the typical gradients of Laplace pressure and viscous stress in eq.~(\ref{NS}). This yields $p=( {\gamma h_0}/{w_0^2}) P=(\eta/(t_0\epsilon^2))P$ which sets $t_0=\eta w_0/(\gamma\epsilon^3)$, where $\gamma$ is the surface tension and $\eta$ is the viscosity. Finally,  estimating the extra-stress tensor in the viscous limit, within the lubrication approximation, leads to:
\begin{eqnarray}
\left( \begin{array}{cc}
\sigma'_{xx} & \sigma'_{xz}  \\
\sigma'_{zx}& \sigma'_{zz}\end{array} \right) =   \frac{\eta}{t_0} \left( \begin{array}{cc}
\Sigma'_{XX} & \displaystyle \frac{\Sigma'_{XZ}}{\epsilon} \\
\displaystyle \frac{\Sigma'_{ZX}}{\epsilon}& \Sigma'_{ZZ} \end{array} \right)  \ .\label{nondim}
\end{eqnarray}
  equation~\eqref{NS} then yields the following set of equations:
\begin{subeqnarray}
\epsilon^2 Re \, \frac{\text{d}{U_X}}{\text{d}T}&=& -\partial_X P+  \epsilon^2 \partial_X \Sigma'_{XX} +\partial_Z \Sigma'_{XZ} \\
\epsilon^4  Re \, \frac{\text{d}{U_Z}}{\text{d}T}&=&   -\partial_Z P + \epsilon^2 \partial_X \Sigma'_{ZX} + \epsilon^2 \partial_Z \Sigma'_{ZZ}\ ,\label{Cz}
\end{subeqnarray}  
where $\displaystyle Re ={\rho v_0 w_0}/{\eta}$ is the Reynolds number.
In the framework of first order lubrication approximation in $\epsilon$, eq.~(\ref{Cz}) simplifies to:
\begin{subeqnarray}
\partial_X P&=&\partial_{Z}\Sigma'_{XZ}\\ \label{Lub}
\partial_Z P &=&0\ .
\end{subeqnarray}
The boundary conditions at the free surface $Z=H(X,T)$ are set by the small-slope approximation of the Laplace pressure and the no-shear stress:
\begin{subequations} \label{BC}
\begin{align}
P\big |_{Z=H}&\,\simeq\,-\partial_X^{\,2} H \label{BC1} \\
\Sigma'_{XZ}\big |_{Z=H}&\,=\,0\ . \label{BC2} 
\end{align}
\end{subequations}
Integrating eq.~(\ref{Lub}) together with eq.~(\ref{BC}) leads to:
\begin{eqnarray}
\Sigma'_{XZ}&=&(H-Z)\,\partial_X^3 H \ .\label{etape1}
\end{eqnarray}
In order to account for viscoelasticity, we use the Maxwell model \cite{Macosko1994,Larson1999}. This simple approach relates the local shear strain rate $\partial_t \varepsilon_{xz}=\partial_zu_x$ and the local shear stress $\sigma'_{xz}$ through:
\begin{eqnarray}
\sigma'_{xz} + \tau \partial_t \sigma'_{xz} &=&\eta \partial_t \varepsilon_{xz} \ ,\label{Maxwell}
\end{eqnarray}
where  $\tau=\eta/G$ is the single characteristic time, with $G$ being the shear elastic elastic modulus. In dimensionless variables, eq.~(\ref{Maxwell}) reads:
\begin{eqnarray}
\Sigma'_{XZ} + \mathcal T\, \partial_T \Sigma'_{XZ} &=& \partial_Z U_X \ ,\label{Maxwellad}
\end{eqnarray}
where $\mathcal T=\tau/t_0$. Substituting eq.~(\ref{etape1}) into eq.~(\ref{Maxwellad}) yields:
\begin{eqnarray}
(1+ \mathcal T \partial_T)\left[(H-Z)\,\partial_X^3 H\right]&=&\partial_Z U_X\ , \label{etape2}
\end{eqnarray}
and spatially integrating eq.~(\ref{etape2}) together with the no-slip boundary condition at the substrate, $U_X \big |_{Z=0}=0$, leads to:
\begin{eqnarray}
(1+ \mathcal T \partial_T)\left[\left(HZ-\frac{Z^2}{2}\right)\,\partial_X^3 H\right]&=&U_X\ .  \label{etape3}
\end{eqnarray}
In order to close the system we invoke mass conservation:
\begin{eqnarray}
\partial_T H +\partial_X Q=0\ ,\label{MassCons}
\end{eqnarray}
where $Q =\int_0^H  U_X\,\text d Z$. Combining eq.~(\ref{etape3}) and eq.~(\ref{MassCons}), together with eq.~(\ref{BC1}),
{finally leads to the governing equation}:
\begin{eqnarray}
\partial_T H&=&-\partial_X\left[ \frac{H^3}{3}\partial_X^{\,3} H\right]\nonumber \\
&&-  \mathcal T\partial_X\left[ \frac{H^2}{2} \partial_T H\partial_X^{\,3} H  + \frac{H^3}{3}  \partial_T\partial_X^{\,3} H  \right] ,\label{VTFE}
\end{eqnarray}
that we will refer to as the viscoelastic thin film equation (VTFE).  Note that by setting $\mathcal T=0$ in eq.~(\ref{VTFE}) one recovers the well know 2D capillary-driven thin film equation for Newtonian fluids \cite{Orchard1961,Oron1997,Craster2009,Blossey2012}. 
{Note also that} eq.~(\ref{VTFE}) is a particular case of {the Jeffreys-based viscoelastic thin film equation \cite{Rauscher2005,Munch2006,Blossey2006,Blossey2012}} for which the early time constant of the Jeffreys model has been set {equal} to zero.

\section{General {linear solution}}
Let us now consider the case in which the surface perturbation is small compared to the overall thickness of the film. One can then write $H(X,T)=1 +\zeta(X,T)$ where $ |\zeta(X,T)|\ll 1$.  In this limit, eq.~(\ref{VTFE}) can be linearised as follows: 
\begin{eqnarray}
\partial_T \zeta&=& - \frac{1}{3} \left[ \partial_X^{\,4}  \zeta +\mathcal T\partial_T\,\partial_X^{\,4} \zeta   \right]\ .\label{LVTFE1}
\end{eqnarray}
In order to get rid of the factor 1/3, we redefine time through  $T  \rightarrow 3T$ and $\mathcal T \rightarrow 3\mathcal T$ so that eq.~(\ref{LVTFE1}) becomes:
\begin{eqnarray}
\mathcal L \,\zeta(X,T)=0 \ ,\label{LVTFE}
\end{eqnarray}
where we have introduced the linear differential operator:
\begin{eqnarray}
\mathcal L&=& \left[  \partial_T\left(1+  \mathcal T \partial_X^{\,4} \right)   +\partial_X^{\,4} \right] \ . 
\end{eqnarray}
{In the following,} eq.~\eqref{LVTFE} {will be referred to as} the linear viscoelastic thin film equation (LVTFE). 
{For a sufficiently regular initial condition $\zeta(X,0)=\zeta_0(X)$, the general solution of eq.~\eqref{LVTFE} is given by the convolution of $\zeta_0$ and the function $\mathcal F$ given by:}
\begin{eqnarray}
\mathcal{F}(X,T)
&=&F(X,T)\,\,+\,\, e^{-T/{\mathcal T}} \,\delta(X)\ , \label{solxt}
\end{eqnarray}
{where:}
\begin{equation}
F(X,T)= \int \frac{\text d K}{2\pi} \left[  \displaystyle \exp{ \left(-\frac{K^4T}{1+\mathcal T K^4 } \right)}-e^{  -{T}/{\mathcal T} } \right]e^{iKX}\  .\label{F}
\end{equation}  
Based on the self-similar behaviours observed in previous studies \cite{Salez2012a,Benzaquen2013}, we let the change of variables:
\begin{figure}[t!]
\begin{center}
\includegraphics[width= 1 \columnwidth]{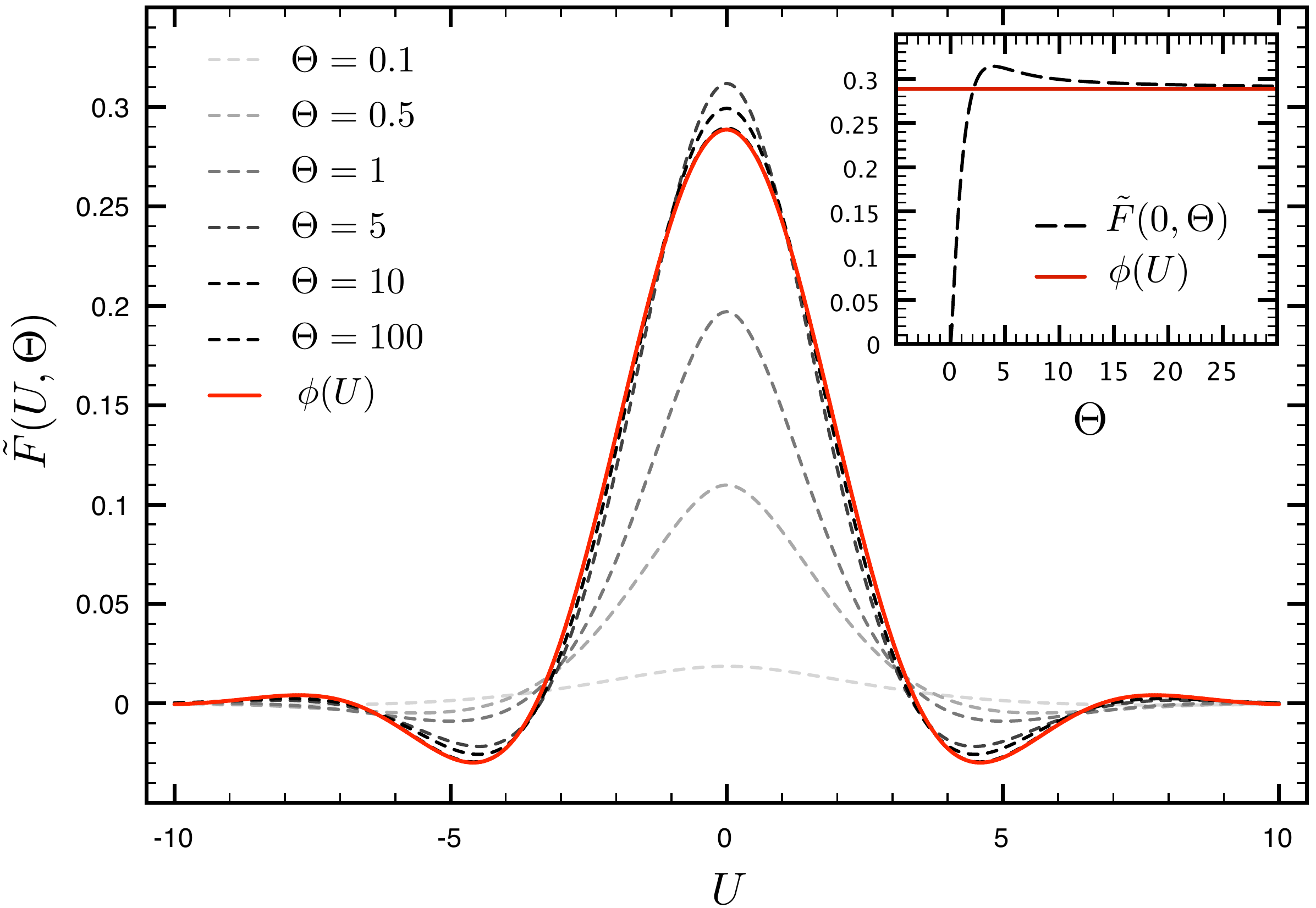}
\end{center}
\caption{Plot of the regular part $\tilde F(U,\Theta)$ {of eq.~\eqref{solxt}},    as a function of $U$, as given by eqs.~(\ref{Gbreve}) and (\ref{soluttheta}), for different rescaled times $\Theta$ (dashed curves). The solid  red  line corresponds to $\phi(U)$ {as defined in eq.~\eqref{GreenLTFE} \cite{Benzaquen2013}.}} 
\label{GreenLVTFE}
\end{figure}
\begin{subeqnarray}
X&=&U\,T^{1/4}\label{CDV1}\\
K&=&Q\,T^{-1/4}\ ,\label{CDV}
\end{subeqnarray}
which, together with eq.~(\ref{solxt}), leads to:
\begin{eqnarray}
T^{1/4} \,\mathcal{\breve{F}}(U,T)&=&\breve{F}(U,T)\,\,+ \,\, e^{-T/{\mathcal T}} \,\delta(U)\ ,   \label{Gbreve}
\end{eqnarray}
where $\mathcal{\breve{F}}(U,T)=\mathcal{F}(X,T)$ and:
\begin{equation}
\breve F(U,T)= \int\frac{ \text d Q}{2\pi} \left[  \displaystyle \exp{ \left(-\frac{Q^4}{1+\frac{\mathcal T}{T} \,Q^4 } \right)}-e^{  -T/{\mathcal T}} \right]e^{iQU} \  .\label{solut}
\end{equation}
Furthermore, defining the rescaled time $\Theta=T/\mathcal T$ we let $\breve F(U,T)=\tilde F(U,\Theta)$ with:
\begin{equation}
\tilde F(U,\Theta)= \int \frac{\text d Q}{2\pi} \left[  \displaystyle \exp{ \left(-\frac{Q^4}{1+\Theta^{-1} \,Q^4 } \right)}-e^{  -\Theta } \right]e^{iQU} \  .\label{soluttheta}
\end{equation}
Figure~\ref{GreenLVTFE} shows a plot of the regular part $\tilde F(U,\Theta)$ {of eq.~\eqref{solxt}} as a function of $U$, as given by eqs.(\ref{Gbreve}) and (\ref{soluttheta}), for different rescaled times $\Theta$ (dashed curves). The solid  red line corresponds to:{
\begin{eqnarray}
\phi(U)&=& \int\frac{ \text d Q}{2\pi} \,e^{-Q^4}e^{iQU}\nonumber \\ 
&=&  T^{1/4} \mathcal{\breve F}_{\text{\,LTFE}}(U,T)  \  ,\label{GreenLTFE}
\end{eqnarray}
where $ \mathcal{\breve F}_{\text{\,LTFE}}$ is the Green's function of the purely viscous linear thin film equation (LTFE) \cite{Benzaquen2013}.}
 The inset shows the central value $\tilde F(0,\Theta)$ as a function of $\Theta$. The function $\tilde F$ seems to converge in time to $\phi(U)$. This convergence can {in fact} be proven rigorously {(see appendix)}. 
Therefore, in the linear case, we recover the intuitive expectation that the longterm evolution of a thin viscoelastic film is well described by a purely viscous thin film equation. The corresponding  convergence time sets the time window in which viscoelastic effects are important and need to be considered when studying the dynamics of the film. We expect this convergence time to scale like $\mathcal T$ and will now study this in more details in the following through a particular example.

\section{Evolution of a Gaussian perturbation}
As mentioned in the previous section, the evolution of the surface displacement $\zeta(X,T)$ for a given initial condition $\zeta(X,0)=\zeta_0(X)$ is given by the convolution of  {the function $\mathcal{F}(X,T)$} and $\zeta_0(X)$:
\begin{eqnarray}
\zeta(X,T)&=& (\mathcal{F} *\zeta_0)(X,T) \nonumber  \\
&=&  (F *\zeta_0)(X,T)+ e^{-T/\mathcal T }\zeta_0(X) \,,\label{Conv}
\end{eqnarray}
according to eqs.~(\ref{solxt}) and (\ref{F}).
Let us consider the case of a normalised Gaussian perturbation of the form:
\begin{eqnarray}
\zeta_0(X)&=&\frac{1}{\sqrt{2\pi A^2}}\,e^{-{X^2}/{(2A^2)}}\ , \label{Gauss}
\end{eqnarray}
where  $a=w_0A$  is the initial horizontal width in {dimensioned} variables. In particular, we focus on  the evolution of the central height $\zeta(0,T)$ of the perturbation for different values of the dimensionless initial width $A$. We let $\zeta(0,T)=\tilde \zeta(0,\Theta)$. Figure~\ref{Delta0} shows a plot of $\log_{10} \tilde \zeta(0,\Theta)$  as a function of $\log_{10} \Theta$  as given by eq.~(\ref{Conv}) together with eq.~(\ref{Gauss}), for different values of  the rescaled dimensionless width $\hat A=A\mathcal T^{-1/4}$, but identical volume. As one can see, regardless of initial width, all the curves are attracted to the same $\Theta^{-1/4}$ grey dashed line at long times, that corresponds to the purely viscous evolution for the considered perturbation volume. Two different situations, that will be discussed in detail in the next part, can be identified in regard to the existence or not of a change in concavity. For the case where concavity changes, the convergence occurs as expected at $\Theta \sim 1$, that is for $T\sim \mathcal T$. In this situation, the Maxwell time thus directly sets the timescale of the crossover between {the} elastic and viscous behaviours.
\begin{figure}[t!]
\begin{center}
\includegraphics[width= 1\columnwidth]{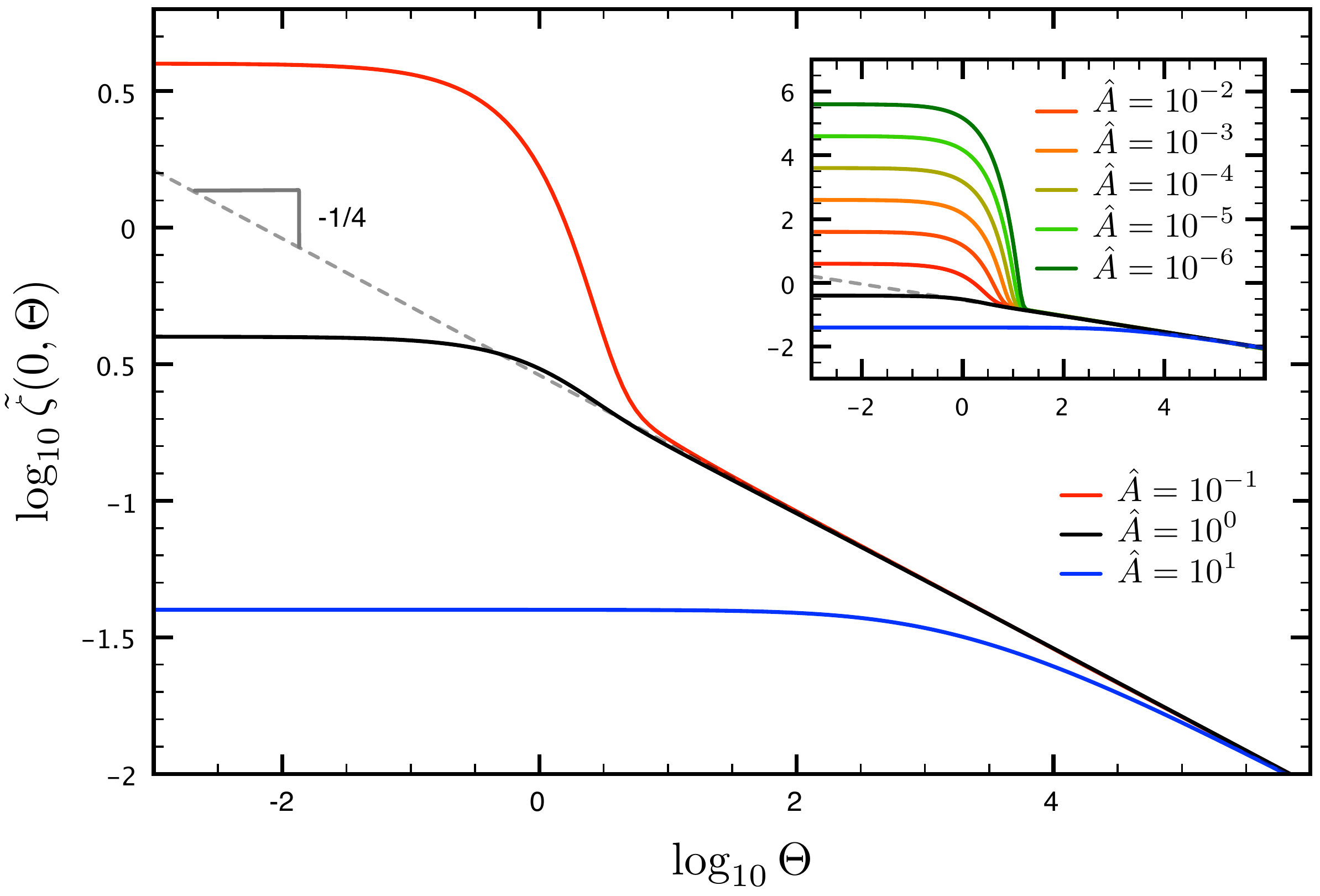}
\end{center}
\caption{Logarithmic amplitude $\log_{10} \tilde \zeta(0,\Theta)$ of the perturbation as a function of logarithmic rescaled time $\log_{10} \Theta$  as given by eqs.~(\ref{Conv}) and~(\ref{Gauss}) for different values of the rescaled dimensionless initial width $\hat A=A\mathcal T^{-1/4}$, but identical volume. The inset shows the same plot for smaller values of $\hat A$. The grey dashed line corresponds to a purely viscous  $\Theta^{-1/4}$ scaling.} 
\label{Delta0}
\end{figure}

\section{Transient levelling exponent}
In a similar way to \cite{Christov2012}, we let the time-dependent levelling  exponent  $\alpha(\Theta)$ be defined as:
\begin{eqnarray}
\tilde\zeta(0,\Theta)&=&\tilde\zeta(0,0)\,\Theta^{-\alpha(\Theta)} \ .
\end{eqnarray}
or equivalently:
\begin{eqnarray}
\alpha(\Theta)&\hat =&-\frac{\log_{10} \tilde\zeta(0,\Theta)-\log_{10} \tilde\zeta(0,0)}{\log_{10} \Theta}  \ .
\end{eqnarray}
The slopes of the curves in Fig.~\ref{Delta0} read:
\begin{eqnarray}
\frac{\text d \log_{10} \tilde\zeta(0,\Theta)}{\text d \log_{10} \Theta}& =& - \alpha(\Theta) - \alpha'(\Theta)\, \Theta \log_{10} \Theta \ , \label{slope}
\end{eqnarray}
which implies that for $\Theta=0$, $\Theta=1$, and $\Theta\rightarrow+\infty$ for which $\alpha'(\Theta)=0$, the slopes of the curves  correspond to the levelling  exponent $\alpha$. Note that this is also relevant for all time windows where $\alpha'(\Theta)\, \Theta \log_{10} \Theta\ll \alpha(\Theta)$, and thus in particular for the crossover region where $\Theta\sim1$. Figure \ref{alpha} shows a plot of the absolute values of the slopes of the curves in  Fig.~\ref{Delta0}  as a function of reduced time $\Theta$, for different rescaled widths $\hat A$ of the initial profile, but identical volume. In Figs. \ref{Delta0} and \ref{alpha}, two different situations can be distinguished. The first one is the situation for which the crossover from $\alpha=0$  to $\alpha=1/4$  happens with no change of concavity in Fig.~\ref{Delta0} or, equivalently, with no overshoot in Fig.~\ref{alpha}  (see blue curve  with $\hat A=10$ on both figures). The second one is the situation for which the aforementioned crossover takes place with a  change of concavity in Fig.~\ref{Delta0} or, equivalently, with an overshoot in Fig.~\ref{alpha} (see curves $\hat A=1$ and below on both figures). To understand the first situation, we recall that for a purely viscous liquid, a given initial perturbation takes a certain time $T_{\text {cv}}$ to converge to the self-similar attractor \cite{Benzaquen2013,Baumchen2013,Backholm2014}. In the particular linear case considered here, this time is expected to scale as the width of the initial perturbation to the power 4. Therefore, the first situation is that for which $T_{\text{cv}}\gg \mathcal T$ and the elastic delay of the evolution has no visible impact as it is shorter than that of the convergence towards the self-similar attractor. On the other hand, the second situation corresponds to $T_{\text{cv}}\ll \mathcal T$ and thus reveals the full viscoelastic behaviour. The transition between the two situations is thus given by $T_{\text{cv}}\sim \mathcal T$, which means $\hat A^4=A^4\mathcal T^{-1}\sim 1$, or in real variables:
\begin{eqnarray}
{a^4}\sim \tau\frac{\gamma h_0^3}{\eta}\sim\frac{\gamma h_0^3}{G}\ .
\end{eqnarray}
Note that one must also keep $h_0\ll a$ for the lubrication approximation to remain valid. Interestingly, the transition between the two situations described above appears to be controlled by the ratio of the width of the perturbation to a new length scale that combines the elastocapillary length and the film thickness. This length scale is simply the length explored by levelling during a Maxwell time. We shall now focus on the second regime, namely $\hat A\ll 1$. To access this regime without breaking the lubrication hypothesis, one can think, for instance, of soft materials with small shear modulus $G$. Here, the clear overshoot in the slope (see Fig.~\ref{alpha}) can give rise to what we will refer to as an anomalous transient levelling exponent. In fact, if an experimental system consisting of a thin viscoelastic film is explored within a time window that is in the vicinity of the maximum of the curve in Fig.~\ref{alpha}, one may read a  levelling exponent $\alpha>1/4$.
This value of the exponent is called \textit{anomalous} since it is not comprised between the elastic (0) and viscous (1/4) standard values. Note that the temporal window of occurrence of this effect may be large due to the logarithmic scale of Fig. 4, and to the proportionality of real times with the time constant $\tau$.  In experimental systems where the viscoelastic timescale $\tau$ is large enough, as with glassy polymer films for instance, one therefore expects to observe an apparently constant and too large levelling exponent that may simply be the result of this anomalous viscoelastic transient overshoot.\begin{figure}[t!]
\begin{center}
\includegraphics[width= 1 \columnwidth]{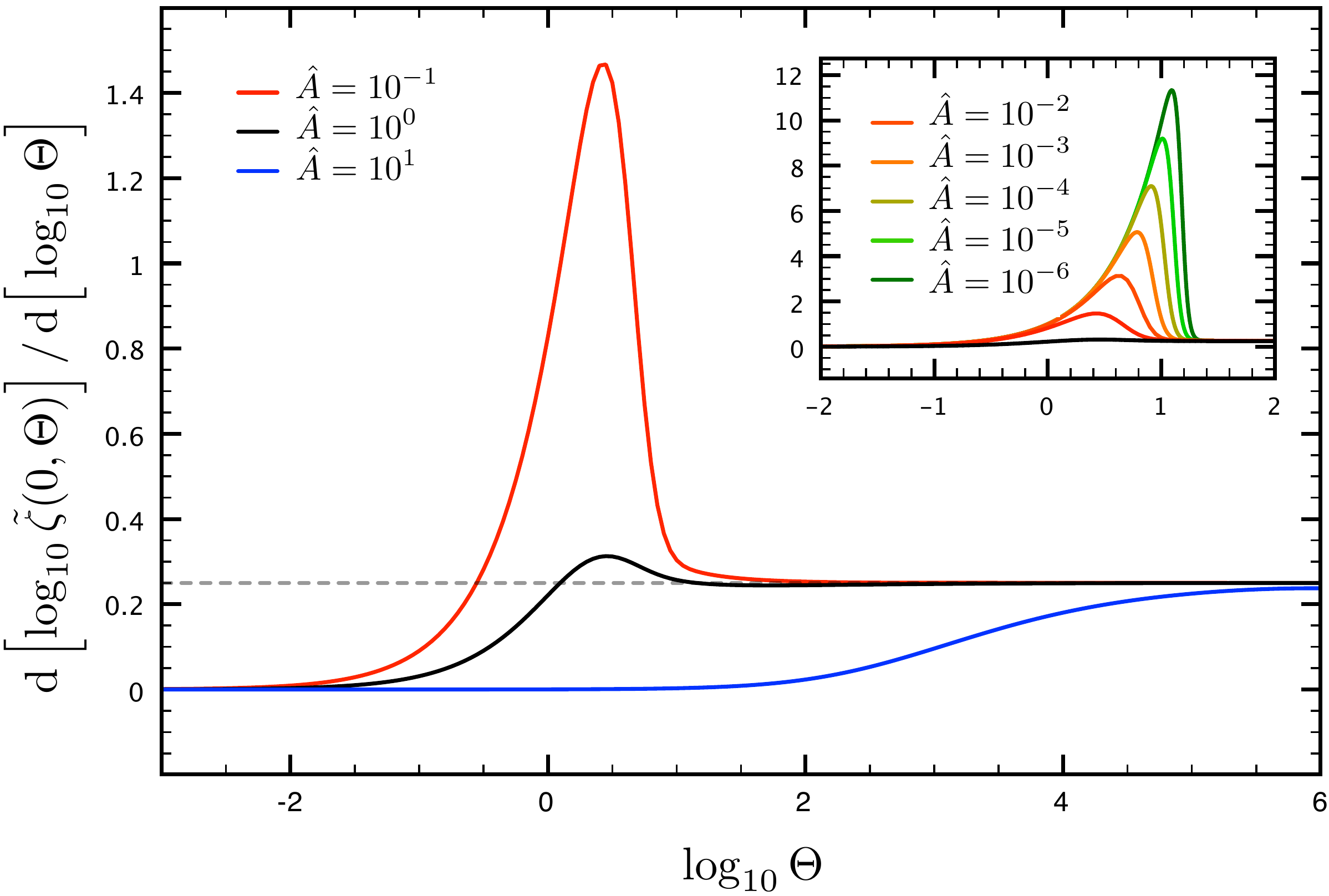}
\end{center}
\caption{Absolute values
 of the slopes of the curves in  Fig.~\ref{Delta0}  as a function of reduced time $\Theta$, for different rescaled dimensionless  widths $\hat A=A\mathcal T^{-1/4}$ of the initial profile (see eq.~(\ref{Gauss})). The inset shows the same plot for smaller values of $\hat A$. The grey dashed line correspond to the purely viscous 1/4 exponent.} \label{alpha}
\end{figure} This result is somewhat analogous to the one highlighted previously by Christov and Stone for diffusion exponents in granular materials \cite{Christov2012}. In their system, the origin of the anomalous scaling exponent is a non-constant diffusion coefficient that changes the structure of the diffusion equation and induces a loss of self-similarity. 

\section{Conclusion}
We have here presented a theoretical analysis of the capillary-driven relaxation of a thin viscoelastic film within the lubrication approximation. After recalling the ingredients of the model, we derived a viscoelastic thin film equation that accounts for viscoelasticity through a Maxwell model. We linearised this equation and  {obtained its general solution}. We proved that the {this solution} converges in time to that of  the purely viscous linear thin film equation, which means that the longterm evolution of a thin viscoelastic film is well described by a purely viscous model. We then looked into the evolution of a  Gaussian initial perturbation and analysed it in terms of its transient levelling exponent.  We  discussed the different situations as a function of the initial width of the perturbation compared to a new length scale, that we identified and that combines the elastocapillary length and the film thickness. For small enough widths, provided that lubrication approximation remains valid, we revealed the possibility of observing anomalous transient levelling exponents if the system is explored within a certain  time window around the characteristic viscoelastic time. This work {should}  be of interest for the study of the levelling dynamics of thin polymer films in the vicinity of the glass transition temperature, for which viscoelastic effects can be important. 

\acknowledgments
The authors wish to thank A. Darmon, J. D. McGraw, K. Dalnoki-Veress and J. A. Forrest for interesting discussions.

{\section{Appendix}
We here rigorously show that the regular part of eq.~\eqref{solxt} as given by eqs~\eqref{Gbreve}, \eqref{solut} and \eqref{soluttheta} converges in time to the Green's function of the purely viscous linear thin film equation (LTFE), as defined in \cite{Benzaquen2013}.     For all $\Theta>0$ and all $Q$ one has:
\begin{eqnarray}
\left| \left[  \displaystyle \exp{ \left(-\frac{Q^4}{1+\Theta^{-1} \,Q^4 } \right)}-e^{  -\Theta } \right]e^{iQU} \right| \,\leq \, g(Q) \ ,
\end{eqnarray}
where $g(Q)\,=\,{2}/({1+Q^2}) $ is a summable function. Therefore, invoking the dominated convergence theorem \cite{Rudin}, one gets:
\begin{eqnarray}
\lim_{\Theta\,\rightarrow \,+\infty} \tilde F(U,\Theta)
&=&\phi(U) \ ,\label{greenltfe1}
\label{greenltfe}
\end{eqnarray}
where $\phi(U)$ is defined in eq.~\eqref{GreenLTFE}.
According to eqs.~\eqref{Gbreve} and \eqref{soluttheta} one may thus write \cite{Benzaquen2013}:
\begin{eqnarray}
\lim_{T\rightarrow\,+\infty}   T^{1/4}\,\mathcal{\breve F}(U,T)&=&T^{1/4}\, \mathcal{\breve F}_{\text{\,LTFE}}(U,T)\ .
\end{eqnarray}}

\end{nolinenumbers}

\end{document}